\documentclass[fleqn,10pt]{wlscirep}
\usepackage{graphicx}
\usepackage{subcaption}
\usepackage{url}

\title{Direct observation of vacuum arc evolution with nanosecond resolution}

\author[1,2]{Zhipeng Zhou}
\author[2]{Andreas Kyritsakis}
\author[1*]{Zhenxing Wang}
\author[1]{Yi Li}
\author[1]{Yingsan Geng}
\author[2,3]{Flyura Djurabekova}

\affil[1]{State Key Laboratory of Electrical Insulation and Power Equipment, Xi’an Jiaotong University, Xi’an 710049, China}
\affil[2]{Helsinki Institute of Physics and Department of Physics, University of Helsinki, P.O. Box 43,
FI-00014 Helsinki, Finland}
\affil[3]{National Research Nuclear University MEPhI, Kashirskoye sh. 31, 115409 Moscow, Russia}
\affil[*]{zxwang@xjtu.edu.cn}

\begin{abstract}
Sufficiently high voltage applied between two metal electrodes, even in ultra high vacuum conditions, results in an inevitable discharge that lights up the entire gap, opening a conductive channel through the vacuum and parasitically consuming large amounts of energy. Despite many efforts to understand the processes that lead to this phenomenon, known as vacuum arc, there is still no consensus regarding the role of each electrode in the evolution of such a momentous process as lightning. Employing a high-speed camera, we capture the entire lightning process step-by-step with a nanosecond resolution and find which of the two electrodes holds the main responsibility for igniting the arc. The light that gradually expands from the positively charged electrode (anode), often is assumed to play the main role in the formation of a vacuum arc. However, both the nanosecond-resolution images of vacuum arc evolution and the corresponding theoretical calculations agree that the conductive channel between the electrodes is built in the form of cathodic plasma long before any significant activity develops in the anode. We show evidently that the anode illumination is weaker and plays a minor role in igniting and maintaining the conductive channel.
\end{abstract}
\begin{document}

\flushbottom
\maketitle
\thispagestyle{empty}

\section*{Introduction}

Electrical breakdowns in general and vacuum electrical breakdowns, in particular, regain their important role in the development of modern technologies. The increasing usage of electric power in different environments inevitably leads to failure of surfaces facing electric fields. In vacuum, even high or ultra-high, the electric breakdowns appear in the form of vacuum arcs. 
The latter in various cases are controllable and serve various technological advances, like ion sources \cite{MEVVA1985} and physical vapour deposition \cite{Anders}. 
However, in most cases the vacuum arcs occur undesirably in an uncontrollable manner, causing problems in various vacuum devices such as fusion reactors \cite{juttner2001cathode,McCracken1980}, vacuum interrupters \cite{slade2007}, satellite systems \cite{de2006multipactor,rozario1994investigation}, X-ray tubes \cite{latham1995} and large particle accelerators. 

Vacuum arcs are particularly detrimental for high precision devices that are built to employ high electric and electromagnetic fields.
Amongst these are multi-kilometre devices such as powerful particle colliders \cite{dobert2005high, clic2016} or tiny micro- or nano-electromechanical system (MEMS or NEMS) and capacitors \cite{lyon2013gap, ducharme2009inside}.
For instance, micro-fabricated devices such as  nano electro-spray thruster arrays for spacecraft are built to withstand large electric fields between electrodes. However, if an arc occurs, the entire chip is destroyed \cite{sterling2013increased}.

Recently, particular attention obtained technologies that employ high accelerating field gradients for high energy physics \cite{clic2016}, free electron lasers \cite{vogel2013results} or medical hadron accelerators for cancer treatment purposes \cite{Chao2009}. Some of these devices are designed to operate with fields up to hundreds of MV/m \cite{clic2016}, which cause intolerably high frequency of vacuum breakdowns. This, in turn, increases wasteful power consumption, reduces the final luminosity of accelerated particles and overall destabilizes the performance of the device \cite{clic2016}.

Vacuum arcs have been under close attention of researchers since the early 1950s. 
In spite of many empirical attempts to describe and quantify the phenomenon \cite{Dyke1953Arc, charbonnier1967electrical, Mesyats2005, latham1995, Anders, fursey2007field}, there is still no consensus on what are the physical processes that lead to its ignition. 
The most common hypothesis is that a vacuum arc starts from micro-protrusions that exist due to different reasons on the metal surface and locally enhance the applied electric field. 
If the local field reaches a critical value (about $10^{10}$ V/m) \cite{Descoeudres2009, Dyke1953I, Dyke1953Arc}, an intense and increasing field emission current appears. 
The latter initiates violent physical processes that within a few ns form a plasma that is able to conduct very high current densities at very small voltage \cite{ArcPIC_1d, arcPIC}, thus rendering the gap conductive.
The high current flowing from the anode to the cathode is accompanied by intermittent light emission from the gap \cite{mazurek1988fast, mazurek1993x}.

Several mechanisms have been proposed to explain the formation of plasma in such high voltage conditions.
Some of them attribute the arc initiation to physical processes appearing in the cathode, while others to ones in the anode side.
For example, Charbonnier et al.\cite{charbonnier1967electrical} suggest that whether an arc is anode-initiated or cathode-initiated depends on the value of local enhancement factor $\beta$ on the cathode. 
Others, make this distinction based on the time delay between the application of a pulse and the occurrence of an arc \cite{yen1984emission,chalmers1982breakdown}. 
Slade \cite{slade2007} proposed to use the gap length as a criterion to consider a vacuum arc to be cathode- or anode-dominated. 
According to the last suggestion, in a short gap electrode systems, the cathode plays a dominant role in initiating breakdowns, while in contrast, for larger gaps, greater than 2 mm, the anode effect takes over the cathode and the processes developed near the electrode with the higher electric potential determine the evolution of the vacuum arc.

Meanwhile, other proposed mechanisms attribute the dominant role for the initiation of a vacuum arc to cathodic processes.
Mesyats et al. \cite{fursey2007field, yen1984emission, mesyats1993ectons, Mesyats_Ecton, mesyats2000cathode} have proposed explosive electron emission mechanism (known as the "ecton" model) on the cathode electrode, which leads to plasma formation. 
Timko et. al. \cite{ArcPIC_1d, arcPIC} reported Particle-In-Cell plasma (PIC) simulations showing that plasma can gradually build up from intensively emitting cathodes due to positive feedback ion bombardment, if a minimum initial neutral evaporation rate is assumed. 
A possible origin for the latter was recently given by Kyritsakis et. al. \cite{kyritsakis2018thermal}, who performed multi-scale atomistic simulations and reported a thermal runaway mechanism on cathodic metal nano-tips. 
All the above proposed mechanisms, along with various experimental studies \cite{Descoeudres2009, Dyke1953Arc, meng2014electrical}, attribute the vacuum arc ignition to processes in the cathode and consider the processes near the anode negligible.
In general, the scientific community has not reached a consensus about the role of each electrode on the vacuum arc ignition.

In the present work we follow the development of a vacuum arc with a nanosecond resolution, for gap lengths varying from 0.5 mm to 5 mm. 
We observe the evolution of the light emission in the gap by an ultra-fast camera, while recording the gap current and voltage simultaneously.
Our experiments in combination with theoretical calculations we conducted, reveal that regardless the gap length, the vacuum arc is always ignited on the cathode side, within a few ns after the field in its vicinity reaches a critical value.

\section*{Results}

\subsection*{Phases of development of a vacuum arc.}

The geometry of our experiments allowed for a clear distinction between the cathode (thin tip) and the anode (flat surface). 
The electrodes were installed in a high-vacuum chamber with a vacuum level of $2.5\times10^{-4}$ Pa, placed at a distance of a few mm from one another. This distance, or the gap length $d_g$, was varied from 0.5 to 5 mm in different experiments. A pulsed high voltage source with a pulse width $\Delta t_V = 1 - 5 \mu$s  was connected to the cathode and provided up to $V_{max} = -40$ kV, which was sufficiently high to ensure the appearance of an arc in every single pulse.

\begin{figure}[htbp]
	\centering
	\includegraphics[width=0.8\linewidth]{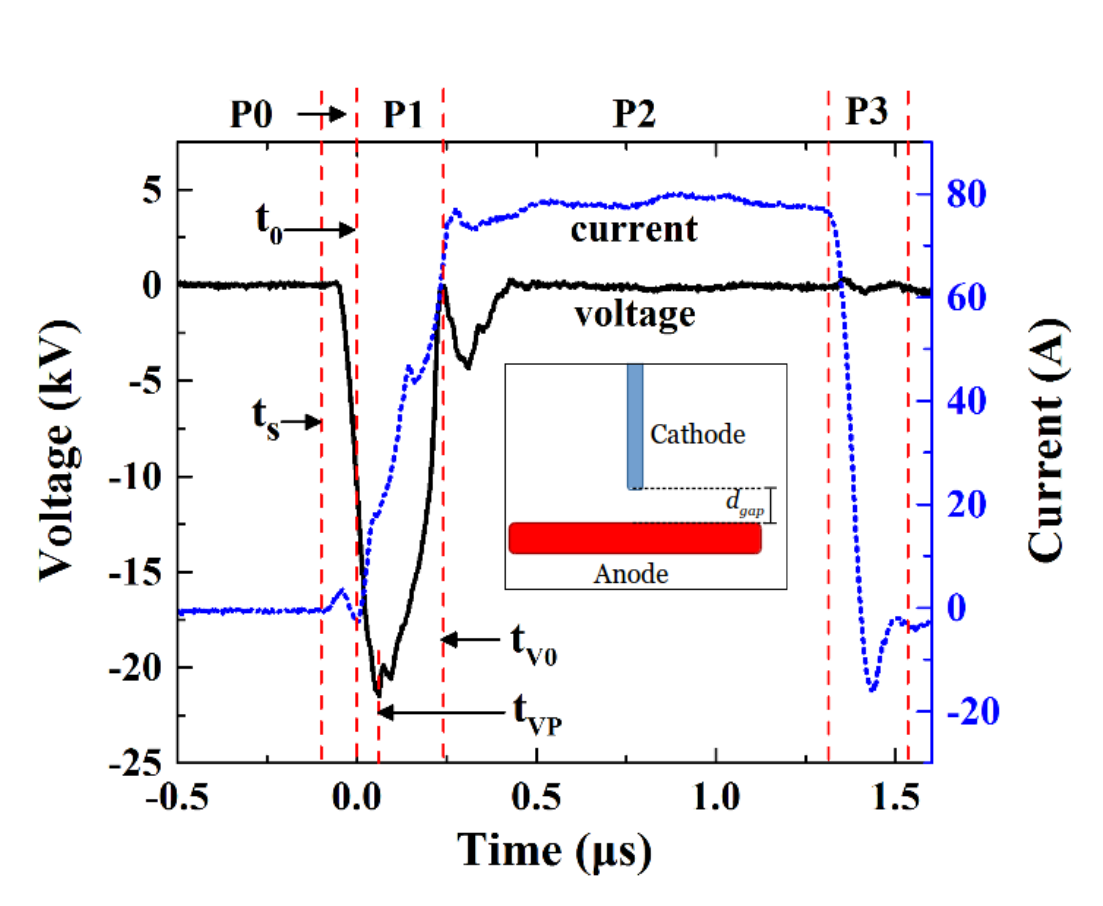}
	\caption{Typical waveforms of the voltage and the current during a vacuum breakdown event registered in the tip-to-plane geometry of the copper electrodes with a gap length of 5mm and a voltage pulse width of 1 $\mu$s. The geometry of the electrodes is shown in the inset. P0-P3 denote different phases of the arc development, $t_s$, $t_0$, $t_{VP}$ and $t_{V0}$ denote the instances when the system started to charge, the current started to rise, the voltage reached its maximum value and the voltage dropped to zero, respectively.}
	\label{fig:phases}
\end{figure}

Fig. \ref{fig:phases} shows typical waveforms of the voltage and current recorded during a breakdown event for the set-up with $d_g$ = 5 mm and $\Delta t_V = 1 \mu$s (see the inset of Fig. \ref{fig:phases} for the geometry of the set-up). 
The abscissa, the left ordinate and the right ordinate show time, gap voltage and gap current, respectively. 
In Fig. \ref{fig:phases}, we identify four main phases of development of a vacuum arc. 
Phase P0, the charging phase, starts when the pulse is applied from the voltage source ($t_s$). 
During P0, the gap capacitor along with parasitic capacitances of the system (a small initial peak in the current waveform) are being charged and the gap voltage starts rising. P0 ends at $t = t_0$, when the current starts rapidly rising up. $t_0$ is also defined as the origin the time axis in our experiments. During the next phase P1, the current rises up to $I_{max} = 80$ A.
We note that the voltage continues growing for a short time until $t_{VP}$; only after this point it drops to a near-zero value, when the current reaches $I_{max}$. 
However, we associate the initial point of the vacuum arc with $t_0$ and not with $t_{VP}$, since the voltage is expected to keep growing after the current through the gap has appeared. 
At this initial stage of the arc, the current is not sufficient to consume the voltage over the gap yet. 
This expectation is corroborated by Simulink \cite{simulink} simulations performed for the same circuit and conditions as used in the experiment (see supplementary material section S1 for the details.)

We also note that the drop of the voltage to a near-zero value and the current rise to $I_{max}$ are completed at approximately the same moment, which we define as $t_{V0}$ and associate with the start of the next phase of the steady arc P2, that lasts until the end of the pulse. The last phase P3 is the discharge decay, during which the voltage and current through the gap drop to zero, completing the vacuum arc process.
We have confirmed the existence of all four phases of the vacuum arc for different voltage pulse widths $\Delta t_V = 1 - 5 \mu s$. 
The corresponding comparison is given in the supplementary material (S3), where we show that longer $\Delta t_V$ only increased the duration of the steady arc phase P2, while the phases P0, P1 and P3, which define the dynamics of arc evolution, are identical and independent of the pulse duration.

\subsection*{Dependence of the waveforms on the gap length} 

\begin{figure}[htbp]
	\centering
	\includegraphics[width=0.7\linewidth]{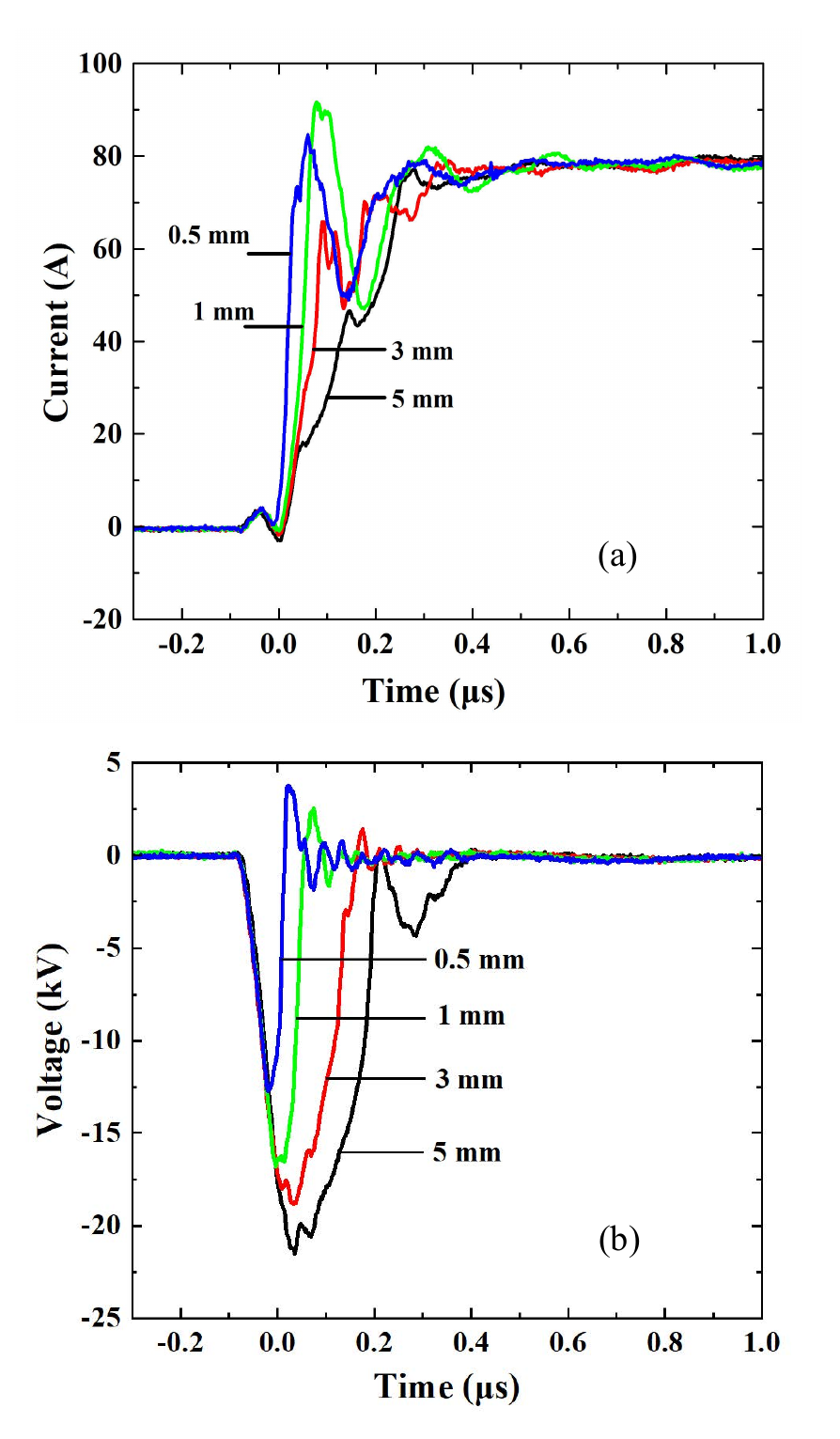}
	\caption{Typical current (a) and voltage (b) waveforms for four different gap lengths, as denoted in the figure. The pulse length is 1 $\mu$s.}
	\label{fig:diff}
\end{figure}

Since the gap length has been suggested to affect the role of the electrodes in the process of vacuum arcing \cite{slade2007}, we performed a series of experiments, where we fixed all the experimental parameters except for $d_g$, varying it from 0.5 mm to 5 mm. 
These results are shown in Fig. \ref{fig:diff}.

\begin{figure}[htbp]
	\includegraphics[width=\linewidth]{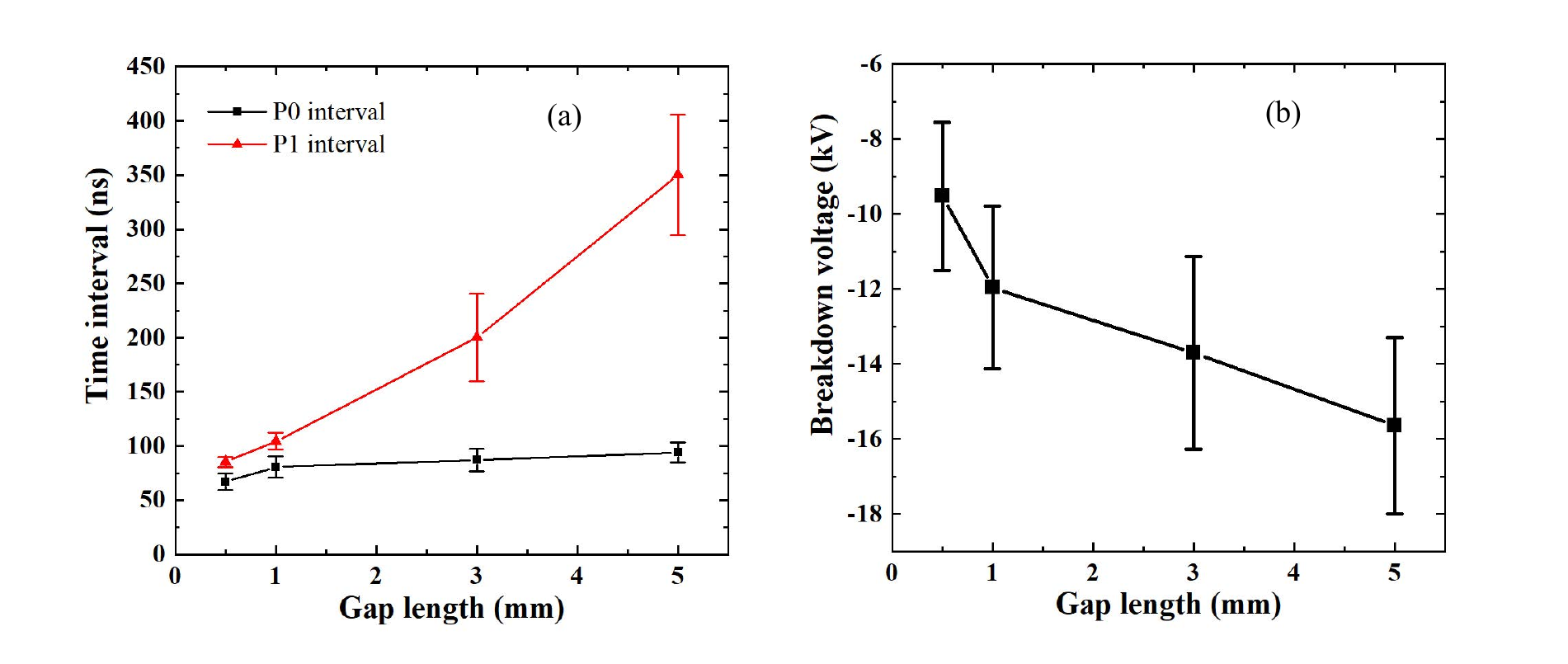}
	\caption{Dependence of the duration of the phases P0 and P1 (a) and the breakdown voltage (b) on the gap length. The corresponding error bars indicate the standard deviation as obtained from 50 measurement repetitions for each gap length.}
	\label{fig:gap_stats}
\end{figure}

As we observe in this figure, both current (Fig.\ref{fig:diff}a) and voltage (Fig.\ref{fig:diff}b) waveforms are affected by the change of $d_g$. 
The shorter this parameter is, the further $t_0$ shifts towards earlier times decreasing the duration of phase P0. 
On the other hand, for longer $d_g$ the current rise phase (P1) lasts longer. 
We analysed these variations and the results are presented in Fig. \ref{fig:gap_stats}(a).
Here the bars show the duration of P0 and P1 phases averaged over 50 independent measurements. 
The corresponding error bars show the standard deviation from the mean value. 
As one can see, the increase of the duration of phase P1 is significant with the increase of $d_g$, while the initial point of the vacuum arc $t_0$ is much less dependent on the size of the gap between the electrodes.

In Fig. \ref{fig:gap_stats}b, the breakdown voltage, i.e. the voltage at $t_0$, is shown to decrease systematically with increasing gap length. 
This clearly indicates that the arc ignites when the local electric field at the apex of the cathode needle reaches a certain critical value.
We calculated the electric field distribution around the apex of the cathode using the finite element method (see the method section) and
found that for all $d_g$, the maximum electric field at $t_0$ is 160$\pm$30 MV/m, which is in surprisingly good agreement with the breakdown fields measurements for flat Cu electrodes \cite{Descoeudres2009, Descoeudres2009_dc}. 

Based on our experiments, we conclude that the increase of the gap length has affected the duration of the identified phases of vacuum arc evolution, however, it did not affect the process of vacuum arcing dramatically, which would have indicated the switch of leading roles of electrodes in this process.

\subsection*{Observation of the vacuum arc development with nanosecond resolution} 
We observed the vacuum arcs through a glass window by an intensified charge-coupled device camera (ICCD, Andor DH334T-18U-04). The electronic gate control of the ICCD allows an exposure time $t_w$ down to 2 ns. However, the physical limitation of the device allows five snapshots per second at maximum. Since the pulse width is only a few $\mu$s, we were able to obtain only one shot per pulse. 

To reproduce the entire evolution of a vacuum arc with a nanosecond resolution, we repeated the experiment numerous times, gradually delaying the moment when the ICCD shot was taken by an interval $\Delta t$ ns with respect to the breakdown time $t_0$ (see the method Section for details).
The repeatability of the experiments is verified and shown in the supplementary material (S2).

In Fig. \ref{fig:main_seq}, we show the full evolution of the light emitted during a vacuum arc for a gap distance $d_g =$3 mm and a pulse length of 5 $\mu$s.
Inspecting the frames in Fig. \ref{fig:main_seq}, we see that a vacuum arc has three major stages with respect to the light emission recorded by the ICCD camera.
During the first stage, which lasts 150 ns (first three frames in Fig. \ref{fig:main_seq}), light is emitted from the tip of the cathode and the anode is dark. 
Also during this stage, the intensity of the light emitted at the cathode gradually increases. 
At 150 ns, the anode begins to radiate and the discharge enters the second stage, characterized by the glow of both electrodes.
During this stage, the anodic glow gradually expands, until it covers the whole gap at 800 ns (10th frame in Fig. \ref{fig:main_seq}). 
Finally, during the last stage, the anodic glow starts decaying even before the end of the voltage pulse and eventually disappears from the gap. 
However, the cathode still glows until 6000 ns after the power supply stops completely at 5000 ns.
After that, the ICCD did not record any radiation from the gap.

In short, based on the analysis of light imaging, we define the three main stages of the vacuum arc development. 
These are the cathode-radiance stage, the anode light expansion stage and, finally, the anode light decaying stage. 
 
\begin{figure}[htbp]
 \includegraphics[width=\linewidth]{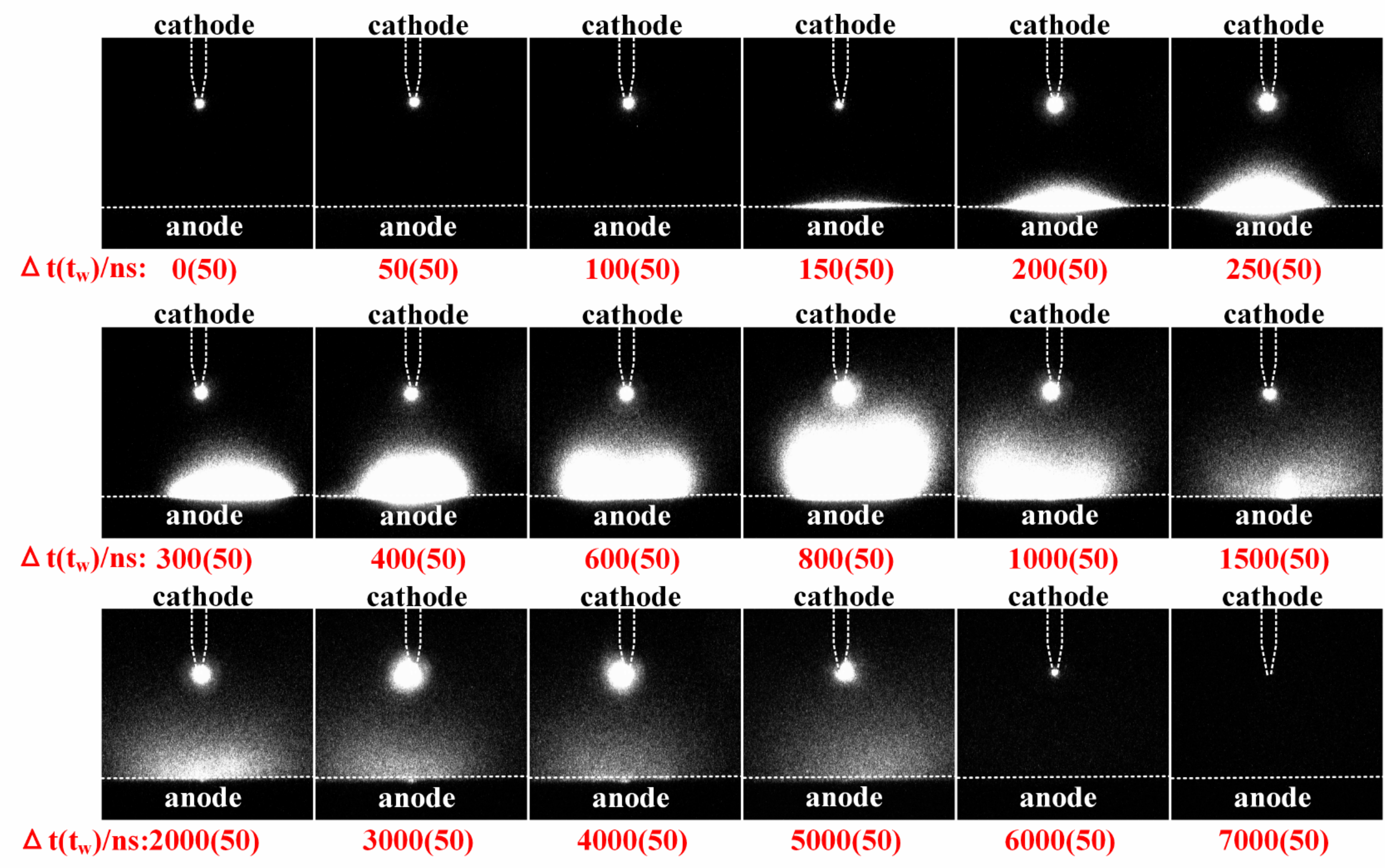}
\caption{Nanosecond-time-resolved light emission of the vacuum arcing process. The gap is 3 mm and the pulse length 5 $\mu$s. The electrodes are outlined by white dashed lines (cathode in the shape of a thin long tip and anode as a large flat surface). The numbers under each frame denote the delay time $\Delta t$. The camera exposure time is $t_w = $50 ns.}
\label{fig:main_seq}
\end{figure}

The time-resolved light emission during the vacuum arcs with gap lengths of 5, 1 and 0.5 mm can be found in the supplementary material (S6).
For different gap lengths we see similar behaviour to the one presented in Fig. \ref{fig:main_seq}, yet with significant differences in the duration of each stage.
The ending points of the three stages are summarized in Table \ref{tab:stages} for all four gap lengths.
The last column of Table \ref{tab:stages} contains the duration of the current rising phase P1, for comparison purposes.
We remind that the end of P1 corresponds to the time when the voltage collapses close to zero and a conductive channel has been formed in the gap.

\begin{table}[htbp]
\centering
\begin{tabular}{|c|c|c|c|c|}
\hline
$d_g$, mm & cathode-radiance, ns & anode light expansion, ns & light decay & current rise (phase P1), ns\\
\hline
5  & 250 & 2050 & 6000 & 350 \\
\hline
3  & 150 & 850 & 6000 &  200 \\
\hline
1 & 40 & 300 & 6000 & 110 \\
\hline
0.5 & 10 & 150 &6000 & 85 \\
\hline
\end{tabular}
\caption{\label{tab:stages} The times when each of the three stages of light emission evolution ended during the development of a vacuum arc. The time is counted from $t_0$ for different gap lengths, $d_g$. The last column corresponds to the end of the current rise phase P1.}
\end{table}

Three important observations emerge from the results of Fig. \ref{fig:main_seq} and Table \ref{tab:stages}.
Firstly, the cathodic radiance appears instantly after the breakdown takes place and the current starts rising. A significant part of the current rise phase P1 coincides with the stage of the cathode radiance. The stage of anode light expansion begins rather late during phase P1 (compare the second and last columns in Table \ref{tab:stages}), i.e. when the gap current is already quite high and the voltage has started collapsing. 
Secondly, the second stage extends far into phase P2 and the moment when the whole gap is bridged by light appears significantly later than the voltage collapse and the formation of a full conductive path in the gap.
Finally, the duration of the two first stages of cathode radiance and anode light expansion depends strongly on the gap length.

\subsection*{Analysis of the cathode and anode light emissions}

The strong flashes of light, which we observe on the snapshots obtained during the vacuum arc (Fig. \ref{fig:main_seq}), do not provide exact information on the intensity of this light. 
To estimate the contribution of each electrode to the glow in the gap, we analyse the intensity of the light emission as follows.

\begin{figure}[htbp]
   \centering
   \includegraphics[width=\linewidth]{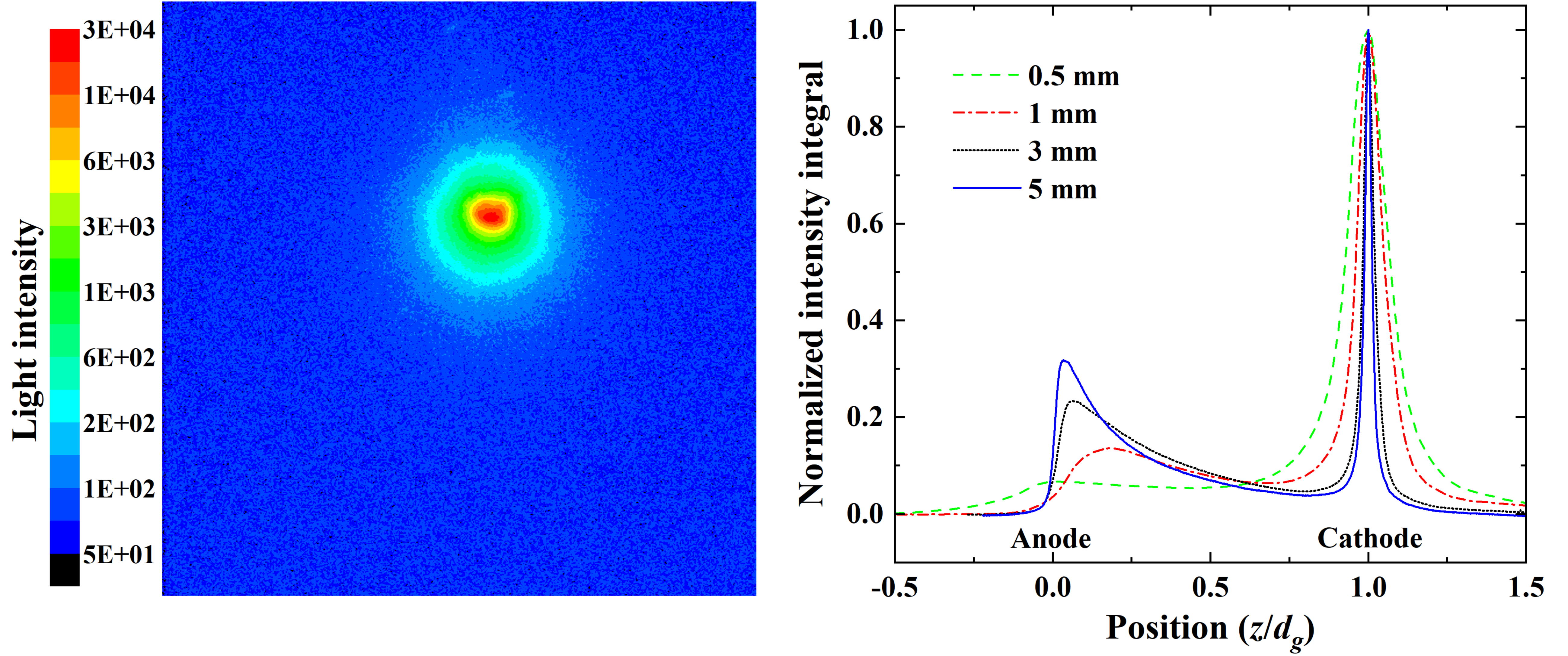}
	\caption{(a) Typical image of the cathode light emission recorded during the ICCD exposure time of 7 $\mu$s, covering the whole pulse duration. (b)Light intensity distribution along the gap. The curves are obtained by summing the intensity in the horizontal direction (parallel to the anode plate), normalizing by its maximum value and averaging 10 different measurement repetitions.}\label{fig:light_emis}
\end{figure}

We first zoom in the camera to focus in the cathode region and set its exposure time to 7 $\mu$s in order to capture the whole discharge process; Fig. \ref{fig:light_emis}(a) demonstrates a typical snapshot of such an exposure.
We see that the light source appears as an extremely focused spherical spot with a maximum intensity at its center that is more than two orders of magnitude higher than the intensity of the surrounding light. The total integrated intensity of this light obtained in the experiments with different  $d_g$ and $\Delta t_V$ did not show dependence on the gap length, but increased linearly with increasing $\Delta t_V$ (see the supplementary material S4 for details).
Furthermore, the full-width-half-maximum (FWHM) range of the peak (i.e. the cathode spot size) is also constant at about 0.1 mm for all gap distances.
Such a consistency of observations indicates that the cathode spot light intensity distribution is stable and constant throughout the whole arc process, regardless the gap length or the pulse duration. 

Zooming out the camera we were able to capture the total intensity of the light emitted during the arc in the whole gap. 
Comparing the intensities of the light emitted at the anode and the cathode, we can examine the contributions of both light sources to deduce a conclusion on which electrode has the leading role in the vacuum arc process. 
In Fig. \ref{fig:light_emis}(b) we plot the normalized intensity distributions (integrated over the lateral directions) along the gap for various gap lengths $d_g$.
Since 
we found that the maximum cathode light intensity is independent of $d_g$, we used it as a reference. 
Hence, all the curves in Fig.\ref{fig:light_emis}(b) are normalized by the peak values of the light intensity at the cathode.
We clearly see that the light intensity at the cathode peak is significantly higher than that at the anode, since both the intensity and the duration of the cathodic glow are significantly higher and longer than those of the anodic one. 
The peak corresponding to the anodic glow, however, is growing with increasing $d_g$. 
It is clear that the anode light appears as a secondary effect caused by the events developed at the cathode.

The fact that the anodic glow begins after the electron current through the gap has risen almost to its maximum value, indicates that the glow at the anode appears as a result of the surface heating by the electron current. 
This scenario is also in line with the fact that the energy available to heat the anode increases with the gap length, since both the duration of phase P1 (current rise) is longer and the breakdown voltage at $t_0$ is higher.

\begin{figure}[htbp]
 \includegraphics[width=\linewidth]{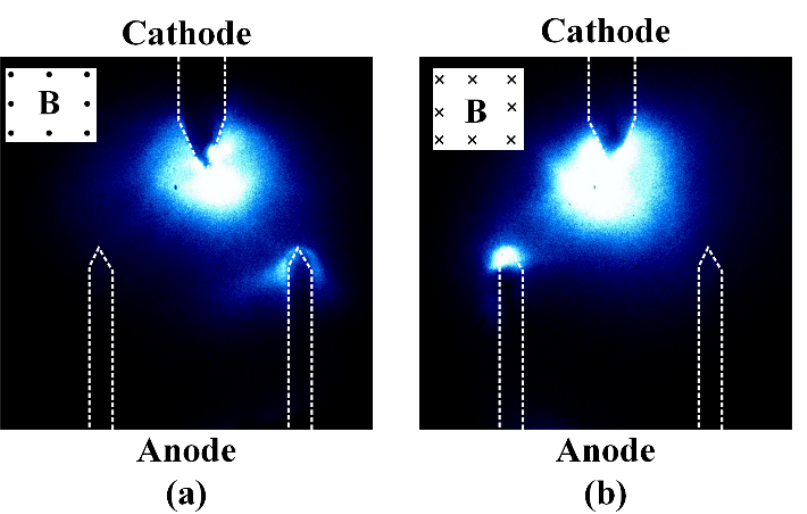}
\caption{Effect of a magnetic field on the discharging process observed for the pulse width $\Delta t_V = 1 \mu$s. Here the anode is shaped as two tips instead of a simple flat plate. The gap length $d_g = 3$ mm is measured between the tops of the tips along the gap. A magnetic field 280 mT in the direction outwards (a) and inwards (b) to the plane of the electrodes was applied. The directions of the magnetic fields are shown in the left top corners of the figures. The exposure time of the camera was 2 $\mu$s and captured the whole discharge process.}
\label{fig:mag_field}
\end{figure}

The above hypothesis regarding the nature of the anodic glow can be confirmed experimentally by examining the response of the anodic glow to the application of a magnetic field perpendicular to the gap current flow. 
For this purpose, we used a different triple-tip configuration of the electrodes. 
A single-tip cathode was placed in the middle of the double-tip anode.
The perpendicular distance from the top of the cathode tip to the tops of the anode tips was $d_g = 3$ mm, and the voltage pulse width was $\Delta t_V = 1 \mu$s. Fig. \ref{fig:mag_field} shows the photographs of the discharges between the electrodes with a magnetic field $B = 280$ mT applied either outwards (Fig. \ref{fig:mag_field}a) or inwards (Fig. \ref{fig:mag_field}b) with respect to the plane of the figure. 

The evolution of the vacuum arc in this experimental configuration was similar to the one we observed for the simple flat plate anode without magnetic field.
However, as can be seen in Fig. \ref{fig:mag_field}, the direction of the field systematically determined where the anodic glow appeared: right when the field is outwards (Fig. \ref{fig:mag_field}a), and left when it is inwards (Fig. \ref{fig:mag_field}b).  
This is consistent with the deflection of negatively charged particles flowing from the cathode to the anode.
This observation confirms that the anodic glow is initiated by electrons impacting at the anode surface and heating it.
The heated anode starts emitting vapour, which interacts with the incoming electrons, producing the glow that expands from the anode surface.

\subsection*{Analysis of the anodic glow}\label{sec:theory}

In the previous sections, we suggested that the anodic glow may start due to the impact of the electrons emitted from the cathode spot.
Here we shall corroborate this explanation by comparing the surface damage of the cathode and anode surfaces and estimating the temperature evolution of the anode.

\begin{figure}[htbp]
 	\includegraphics[width=\linewidth]{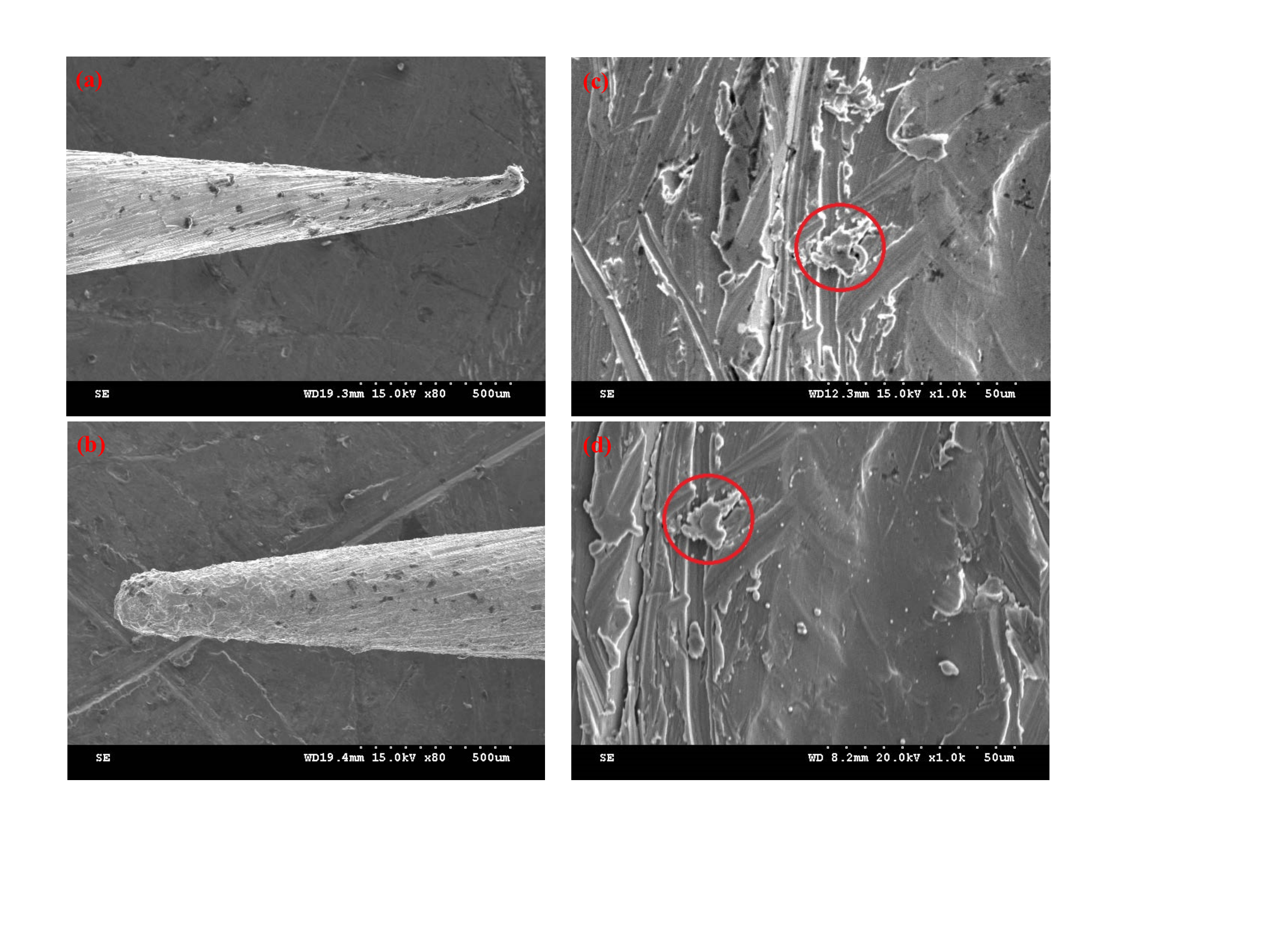}
	\caption{SEM figures for the cathode and anode surfaces both before and after vacuum discharges in a 3 mm gap. (a) cathode surface before discharges; 	(b) cathode surface after discharges; (c) anode surface before discharges; (d) anode surface after discharges. Red circles in (c) and (d) indicate the same 	position. Length scale is indicated by the ruler consisting of 11 little dots at the right bottom corner.}
	\label{fig:SEM}
\end{figure}

In order to assess the degree of surface damage corresponding to the cathodic and anodic glow, we conducted SEM (Scanning Electron Microscope) tests on the cathode and anode surface both before and after electrical discharges. was used. 
Fig. \ref{fig:SEM} shows the corresponding images obtained from a Hitachi S-3000N SEM, for a 3 mm gap. 
We observe clear melting on the cathode surface, while the anode does not show any indications of a melting process.
Many microscopic features found before the breakdown (see, for instance, red circle in Fig.8c) are still present after the experiment (Fig. 8d).
On the contrary, the surface of the cathode is heavily damaged, appearing as a solidified liquid (compare Fig. 8a and 8b). 
The damage of the cathode surface suggests that the corresponding intense glow can be explained by the presence of a fully-developed arc plasma \cite{djurabekova2012crater, timko2010mechanism}, while the nearly unchanged anode surface needs to be examined further by estimating its temperature.

To this end, we solved numerically the one-dimensional time-dependent heat diffusion equation for a flat copper plate, as described in the method section. 
Knowing both waveforms of the voltage over the gap and the current through the gap, we can estimate the heating power deposited by the electrons arriving at the surface of the anode (see the method section for details). This estimation is done under the assumption that during the cathode-radiance stage of the arc (before the anode starts glowing), the electrons are freely accelerated by the gap voltage and deposit all their energy on the anode plate.

\begin{figure}[htbp]
 \centering
 \includegraphics[width=0.8\linewidth]{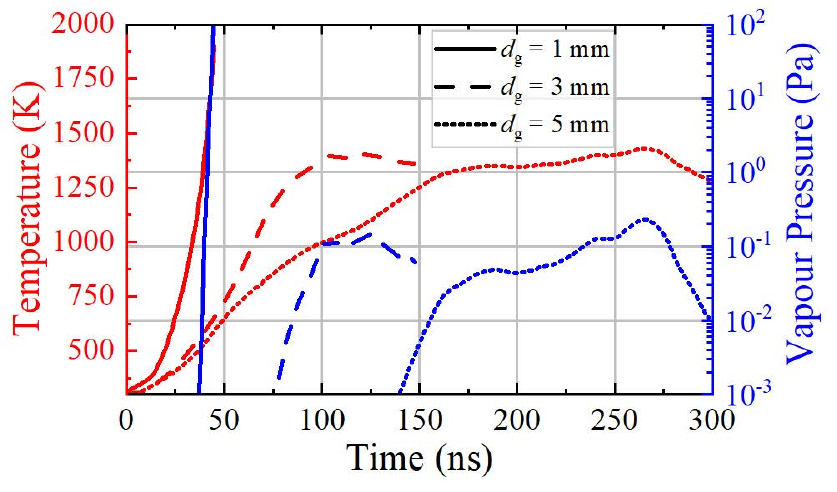}
\caption{Evolution of the temperature at the anode plate (red, left axis) and the corresponding Cu vapour pressure (blue, right axis), calculated for three typical experiments with $d_g = 1$ mm (solid lines) and $d_g = 3$ mm (dashed lines) and $d_g = 5$ mm (dot lines). The maximum depth of the molten region does not exceed 0.5$\mu$m for all cases.}
\label{fig:anode_temp}
\end{figure}

In Fig. \ref{fig:anode_temp} we plot the evolution of the calculated surface temperature of the anode plate and the vapour pressure corresponding to this temperature, during the cathode-radiance stage, for three typical experiments at gap lengths $d_g = 1$, 3 and 5 mm.
We see that the temperatures reach the melting temperature of Cu (1356 K) for all gap distances, with the corresponding vapour pressure exceeding 0.1 Pa. 
Such a pressure corresponds to neutral atom densities of the order of $10^{18}$ -- $10^{19}$ m$^{-3}$.
The electrons colliding with these neutral atoms cause the expansion of the glow that appears near the anode.

Furthermore, the vapour pressure reaches this range in a different time scale for each gap length (at 40, 100 and 230 ns for $d_g = $ 1,3,5 mm correspondingly).
These time scales are in agreement with the times that the anodic glow appears (see Fig. \ref{fig:light_emis} and Fig. S7-S8 of the supplementary material), i.e. the duration of cathode-radiance stage (see Table \ref{tab:stages}).
The latter increases for increasing gap distance, because the electron beam spreads broader reducing the deposited heat per unit area. 
This means that with increase of the gap length the anode needs longer time to be heated to the temperature sufficient for intensive evaporation.

Finally, although the electron-deposited heat is sufficient to melt the anode surface for all cases, the maximum depth of the molten region does not exceed 0.5$\mu$m for all cases. Therefore, the heat does not penetrate in a depth that is sufficient to cause a noticeable melting damage, as shown in the SEM images of figure \ref{fig:SEM}.

\section*{Discussion}

The simultaneous analysis of the measured voltage-current waveforms and the light images at a nanosecond resolution provides deep a insight into the evolution of the vacuum arc process. 
As we saw in the results section, the vacuum arc always ignites at the voltage resulting in a local electric field near the cathode of $\sim$ 160 MV/m. At this point, a measurable electron current through the gap starts rising. 
Instantly after this moment, a spherical-shaped, localized and dense glow appears near the cathode tip.
The latter gradually expands as the current rises towards the external-circuit-limited value of about 80 A, while the gap voltage gradually collapses.
Before the voltage collapses completely, another glow appears in the anode region, slowly expanding and covering eventually the whole gap. 
However, shortly after the gap is bridged by light, the anodic glow starts decaying, although the arc continues burning in a stable high-current low-voltage regime. At this point, only the intense light from the cathode remains, maintaining the arc. The cathode light decays slowly after the voltage pulse stops fully.

The cathodic glow may be explained by either a strong temperature rise due to intensive electron emission, or by the development of a local vacuum arc (i.e. plasma) near the cathode. 
If the first scenario is true, the tip is heated to a very high temperature due to the Joule and Nottingham effects and emit light due to black body radiation. 
It is well known that any type of intensive electron emission (field, thermionic or mixed) is limited by the space charge effect.
For very high emitted current density, the forming cloud of the emitted electrons creates a significant space charge in the vacuum above the emitting surface, that screens the applied field and thus causes a negative feedback that reduces the emission.
The maximum limit of the current density that can be emitted from a cathode depends on the local surface field and the total applied voltage. 
This dependence is given by the Child-Langmuir law \cite{Child_SC}. 
Using this law, we estimated the maximum emission current limited by the space charge for the geometry of the current experiments.
After calculating the distribution of the local electric field around the cathode surface for a given voltage by the finite element method  (see Methods section), we integrated over the whole cathode surface the current density obtained by the Child-Langmuir law.
The details of this calculation can be found in the supplementary material (S5).

Comparing the experimental current waveform with the calculated space charge limit, we can verify the nature of the cathodic glow.  
Shortly after the beginning of the current rise phase P1 (see Fig. \ref{fig:phases}) and the cathodic glow stage, the measured current through the gap exceeds significantly the value limited by the space charge. (see Fig. S5 in the supplementary material).
Furthermore, this happens long before an anodic glow begins in all experiments with different $d_g$.
This consideration allows us to conclude that the strong light emission seen at the cathode surface cannot be caused by electron emission heating phenomena, since the currents measured through the gap at the time of the strong cathodic glow are much higher than those limited by the intensively building-up space charge. 
Furthermore, the spherical shape and the high intensity of the cathodic glow that are independent of the gap length, also confirm that the glow is due to the full arc plasma that forms within a few ns after the local field reaches a critical value.

The anodic glow appears to be rather different by nature from the cathodic one. 
The response of the system to the application of a magnetic field shown in Fig. \ref{fig:mag_field} indicates that the anode glow appears as a result of impacts of electrons on the anode surface.
The electrons accelerated from the cathode deposit their energy on the anode surface. 
This energy can only be transformed into heat, which causes high temperatures and significant metal vapour, that gradually expands to fill the whole gap.
The collisions of the vaporized neutral Cu atoms with the electron beam causes the apparent anodic glow, while they hinder the further heat deposition on the anode, thus self-regulating its temperature.

After the conductive channel is formed, the energy available to heat the anode decreases rapidly due to the collapse of the gap voltage.
As a result, the anode material gradually cools and stops providing Cu vapour. 
Thus, the vapour gradually expands and diffuses away, leading to the decay of the light radiation on the anode surface and in the gap.
On the contrary, the cathode radiance spot remains at the same high intensity until the end of the pulse.
This scenario is confirmed by our anode heat calculations, which show that the electron impact power available in the gap is sufficient to heat the anode to high temperatures, within time intervals that are in agreement with the ICCD camera measurements.

Given the above, although we do not investigate here whether the anodic glow fulfils the plasma criteria, we can conclude that in contrast to the cathodic glow, it is neither stable nor necessary to sustain the arc.
It is rather a transient side-effect of the fact that the gap voltage does not collapse immediately after the arc is ignited, but has a delay time that depends on the gap length.

In summary, by conducting vacuum breakdown experiments between Cu electrodes under rectangular pulse voltages and using a high-speed ICCD camera, we reconstructed the entire vacuum arc process with nanosecond resolution. 
Combining these results with theoretical estimations of the electron emission characteristics, breakdown currents and anode heat evolution, we conclude:

\begin{enumerate}
	\item The vacuum breakdown is triggered at the cathode once the surface field reaches a critical value of about 160 MV/m. 
	Immediately after this, a localized intense radiance appears near the cathode; strong experimental and theoretical evidence shows that the latter is produced by a dense plasma that is formed at the cathode and drives the discharge allowing the gap current to grow to breakdown values.
	\item A while after the breakdown initiation, another light emission starts from the anode, gradually growing to cover the whole gap.
	We suggest that this glow results from the electrons that escape the cathodic arc plasma and bombard the anode. 
	Our heat diffusion calculations and the observations of the deflection of the anode glow in a magnetic field, confirm the correlation between the anodic glow and electrons escaping from the cathode plasma.
	\item Although both the cathode and the anode contribute in the vacuum arc evolution, the role of the cathode is more crucial, since the processes developing at the cathode surface initiate the breakdown and maintain constant radiance from the cathode surface throughout the whole arc process, driving the arc in a stable manner.
	On the contrary, the anode is active only during a fraction of the arc process, and the anode glow covers the gap long after a full conductive channel is established.
\end{enumerate}

\section*{Methods} \label{sec:method}

\subsection*{Experimental Set-up}

\begin{figure}[htbp]
 \includegraphics[width=\linewidth]{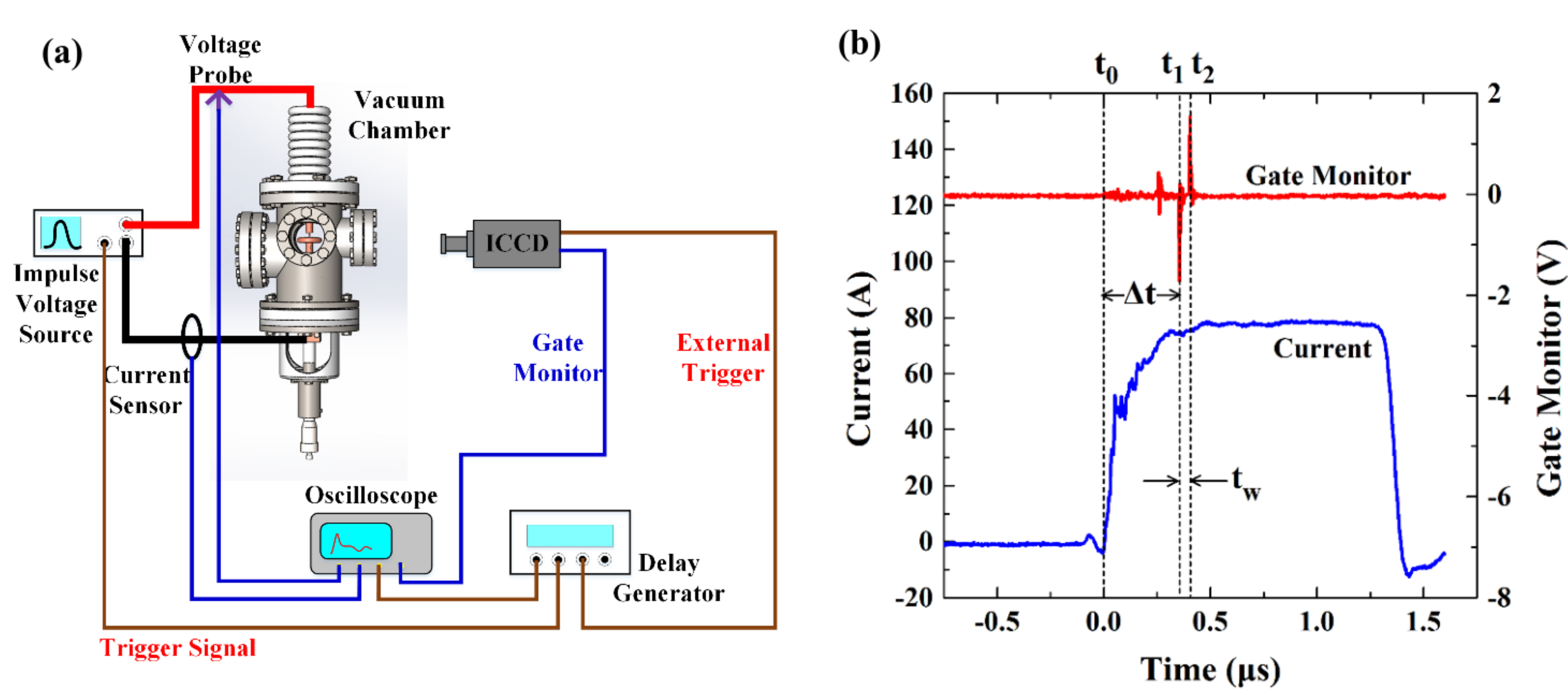}
\caption{(a) Schematic of the experimental set-up. (b)Timing set-up diagram. $t_0$: start point of current rising; $t_1$: opening the ICCD shutter; $t_2$: closing the ICCD shutter; $t_w=t_2 - t_1$: exposure time; $\Delta t = t_1 - t_0$: the beginning of an image capture.}
\label{fig:setup}
\end{figure}

Figure \ref{fig:setup}(a) is the schematic diagram of the experimental set-up.
Electrical discharges were triggered in a demountable stainless steel chamber that was pumped to a pressure of $2.5 \times 10^{-4}$ Pa by a turbo molecular pump. 
A pair of electrodes was installed in the chamber and the gap length was adjusted by a micrometer manipulator.
High voltage was provided by the pulsed voltage source with the output voltage set to -40 kV and a discharge current with the maximum value of 80 A, which was determined by a 500 $\Omega$ current limiting resistor installed in the circuit.
In addition, the width of the voltage pulse was also adjustable for specific purposes, between 1 $\mu$s and 5 $\mu$s.
The upper and lower electrodes were connected to the high voltage terminal and the ground, respectively.

We observed the vacuum breakdowns through a glass window by an intensified charge-coupled device camera (ICCD, Andor DH334T-18U-04).
This ICCD has an electronic gate control to ensure a minimum exposure time of 2 ns and offers a gate monitor signal to indicate the time instant of an observation.
A high voltage probe (NorthStar PVM-7) was used to measure the voltage across the gap, which has a bandwidth of 110 MHz.
The current through the circuit was measured by a Pearson current sensor (Model 6595) with a bandwidth of 200 MHz.
The voltage signal, the current signal and the gate monitor signal were all recorded by a four-channel oscilloscope.

In addition, a digital delay generator (SRS DG645) controlled the sequence of the experiments, in the manner that is shown in Fig. \ref{fig:setup}(b).
After the beginning of the breakdown at $t_0$, the delay generator causes a delay time $\Delta t$, after which at $t_1$ a signal is given to the ICCD camera shutter to open for an exposure time $t_w$. 

\subsection*{Finite element calculation of the field distribution}

\begin{figure}[htbp]
\centering
 \includegraphics[width=.5\linewidth]{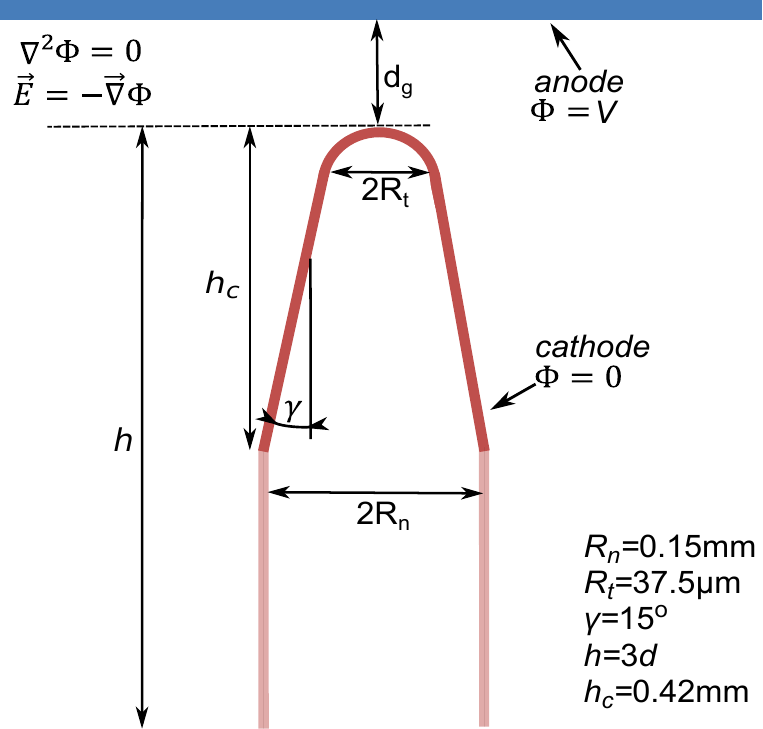}
\caption{Schematic of the finite element simulation geometry.}
\label{fig:schematic}
\end{figure}

We calculated the electric field distribution around the needle by the Finite Element Method (FEM), using the open-source tools Gmsh-GetDP \cite{GetDP}.
The schematic in Fig. \ref{fig:schematic}(a) illustrates the simulated geometry, the equations and the corresponding boundary conditions.
The Laplace equation is solved in the gap, with Dirichlet boundaries at the cathode and the anode.
The cathode tip is simulated as a hemisphere on a cone, which is terminated by a cylinder.
The radii and the aperture angle were chosen based on the geometry of the cathode tips, such as the one shown in the Scanning Electron Microscope (SEM) image in Fig. \ref{fig:SEM}(a).
The total height $h = 3 d_g$ is converged so that its increase does not affect the field on the conical area.

\subsection*{Solution of the heat diffusion equation with the finite difference method}

Since the depth of the heated volume is much smaller than its lateral dimensions, we ignore the lateral heat flow and solve the heat equation in one dimension in order to obtain the heat distribution and evolution. The heat equation can be written as
\begin{equation} \label{eq:heat}
	C_v \frac{\partial T}{\partial t} = \frac{\partial T}{ \partial z} \left[ \kappa(T) \frac{\partial T}{\partial z} \right] + p(z)
\end{equation}
where $C_v$ is the volumetric heat capacity, $\kappa(T)$ is the heat conductivity of Cu (depends on the temperature) and $p(z)$ is the deposited heat density as a function of the material depth $z$.
 
For the purposes of this work we approximate $p(z)$ with the “continuous slowing down approximation - CSDA” \cite{estar}.
This means that we consider the deposited heating power density to be constant over the CSDA range $z_d$ and zero deeper than it, i.e. $p(z > z_d) = 0$ and $p(z<z_d) = P/z_d$, with $P$ being the deposited heating power per unit area.
The heat conductivity of Cu is calculated by applying the Wiedemann-Franz law on the values of the Cu electric conductivity found in the literature \cite{Matula1979, Gathers1983} and the CSDA range $z_d$ is found by the ESTAR database \cite{estar}.

In order to estimate the deposited heating power $P$, we have to consider a certain distribution for the current density of the electron beam impinging on the anode. 
We assume that this roughly follows a Gaussian distribution with its width $\sigma$ being estimated from the width of the anodic glow appearing in the camera images as $\sigma =$ 0.3, 0.85, and 0.95 mm for $d_g = $ 1, 3, and 5 mm correspondingly. 
Then the peak power in the center of the beam can be found as $P(t) = V(t)I(t) / (2 \pi \sigma^2)$, where the product $V(t) I(t)$ gives the total power deposited on the anode by the discharge and is taken from the measured waveforms.

With the above assumptions, we solve the equation (\ref{eq:heat}) to obtain the evolution of the temperature depth profile evolution $T(z,t)$.
The equation is solved over a total depth domain of 20 $\mu m$, sampled at 512 equidistant points, with zero-heat-flux Neumann boundary conditions at the boundaries.
The initial temperature is assumed to be uniform at 300K and we used a forward-time-central-space finite difference integration scheme \cite{orlande2017finite} with a time-step of 0.2 ps. 
When the temperature profile $T(z,t)$ is calculated, the corresponding vapour pressure and vapour density are obtained from the surface temperature $T(0,t)$ by interpolating tabulated temperature-pressure data \cite{vapor_data, alcock1984vapour, safarian2013vacuum}.

\section*{Acknowledgements}

Y. Geng was supported by National Key Basic Research Program of China (973 Program) (No. 2015CB251002). Z. Wang was supported by the National Natural Science Foundation of China (No. 51807147). A. Kyritsakis was supported by the CERN K-contract (No. 47207461). F.\;Djurabekova acknowledges gratefully the financial support of the Academy of Finland (Grant No. 269696). Z. Zhou was supported by the project of China Scholarship Council (No. 201806280259).

\section*{Author contributions statement}

Y.G planned the project and supervised the research. Z.Z and Z.W. designed and conducted the experiments and data analysis, as well as proposed and developed the mechanism hypothesis, together with A.K. and F.D. A.K. conducted the accompanying calculations and interpreted their connection to the experiments together with F.D. Y.L. conducted the experiments together with Z.Z. and Z.W. Z.Z. wrote the manuscript together with A.K. All authors reviewed the manuscript. 

\section*{Additional information}
\textbf{Competing interests:} The authors declare no competing interests. \\

\section*{Data availability}
The data sets generated and analysed during the current study are available from the corresponding author on reasonable request.

\appendix

\renewcommand{\thesubsection}{S\arabic{subsection}}
\renewcommand{\thefigure}{S\arabic{figure}}

\section*{Supplementary material}

\subsection{Determination of the breakdown instant}

\begin{figure}[htbp]
	\centering
	\includegraphics[width=0.8\linewidth]{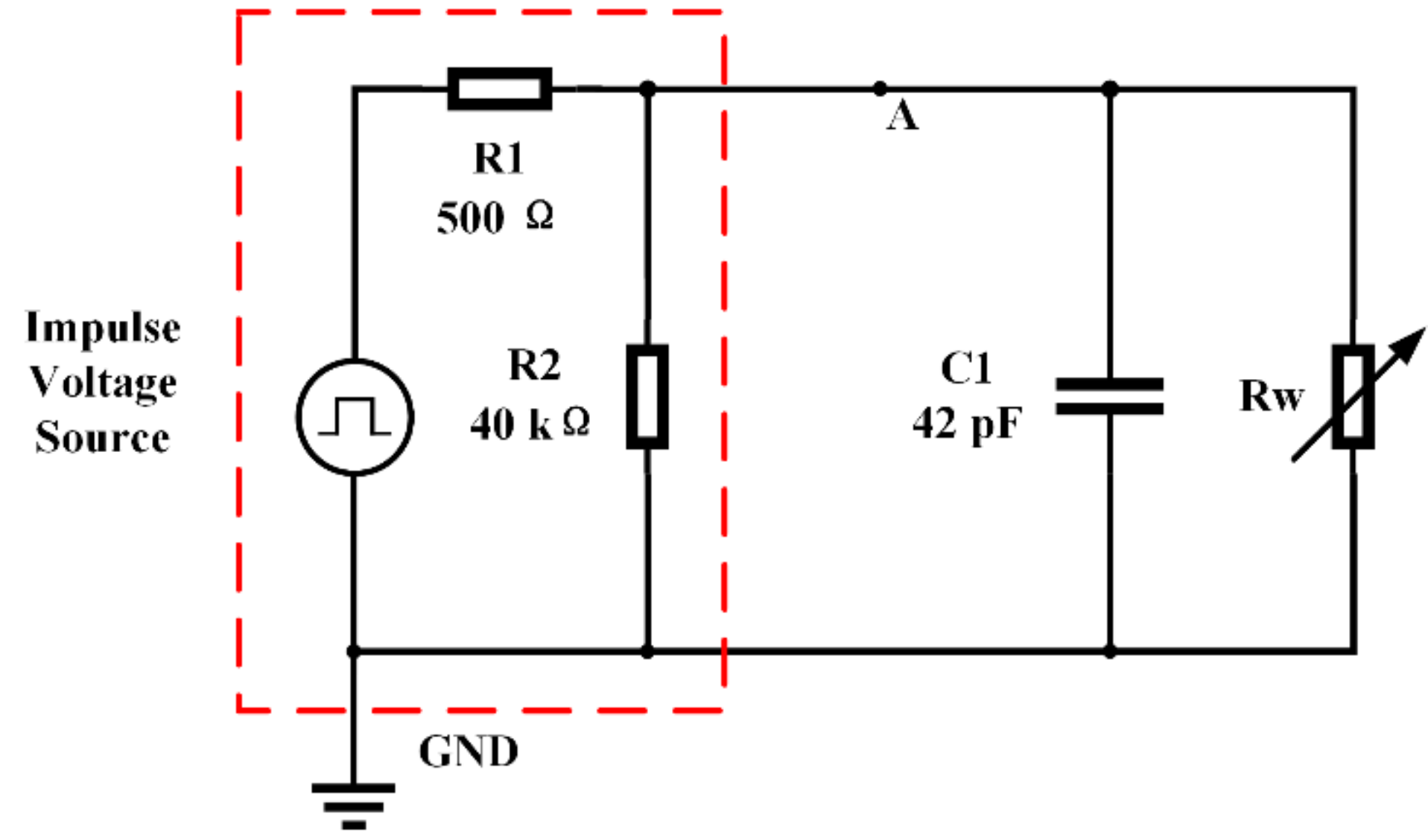}
	\caption{Schematic of the simplified circuit representing the vacuum arc system.}
	\label{fig:circuit}
\end{figure}

Figure \ref{fig:circuit} demonstrates a simplified circuit model of the experiments. 
The impulse voltage source consists of three components enclosed by a dash line rectangular, an ideal pulse voltage source, a current-limiting resistor R1 and a discharge resistor R2. R1 is 500 $\Omega$, and R2 is 40 k$\Omega$. 
The vacuum gap together with the chamber are represented by a capacitor and a variable resistor in parallel. 
The total capacitance of the system C1 was measured 42 pF.
Rw varies from infinity before the breakdown (open circuit) to a very small value after the full breakdown to (short-circuit).
The total current is measured using the current sensor on the wire between the high voltage terminal of the voltage source and the upper connector (connected to the cathode) of the chamber, i.e. at the point $A$ shown in the schematic. 
The gap voltage $V$ is measured at the same point, between the gap electrodes.
Finally, the gap current is obtained by subtracting the capacitive component $C_1 dV / dt$ of the current flowing through $C_1$ from the total current as measured at $A$.

\begin{figure}[htbp]
	\centering
	\includegraphics[width=\linewidth]{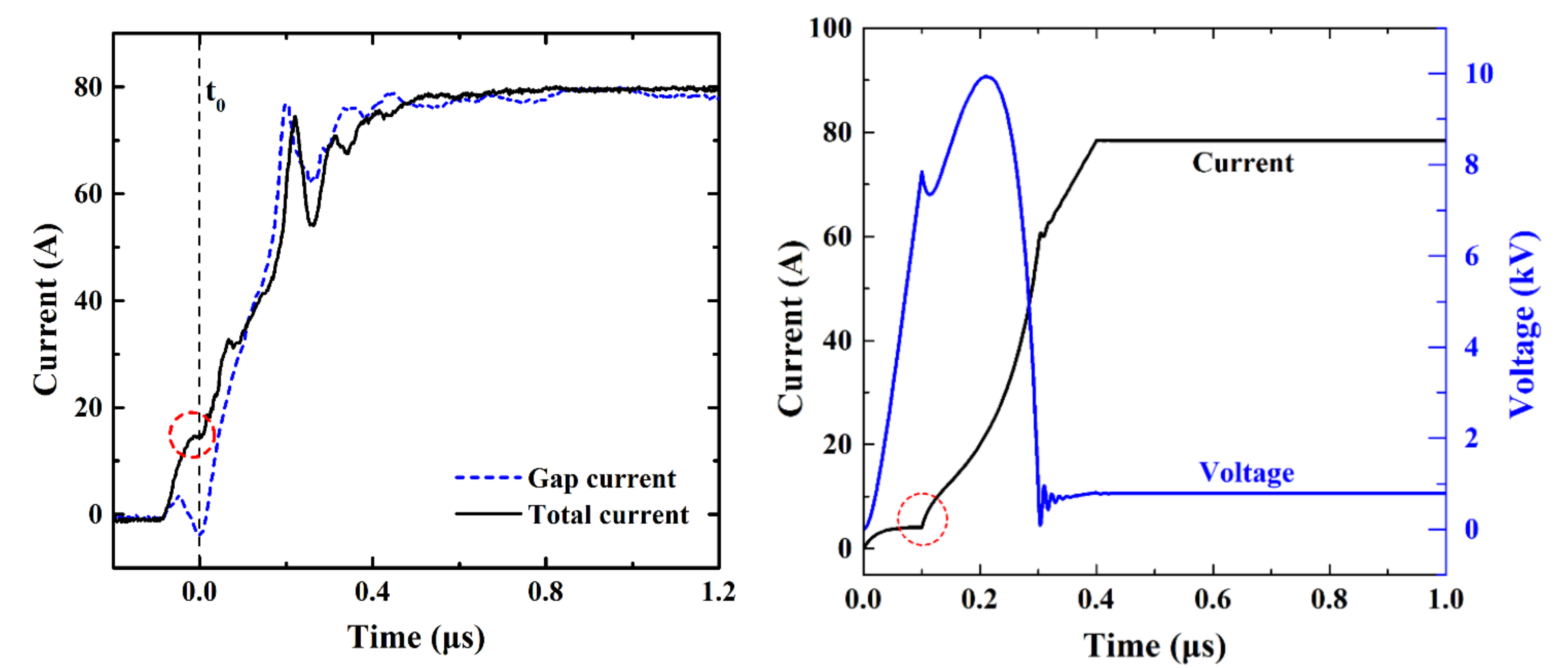}
	\caption{(a) Waveforms of the total current and the gap current during a vacuum breakdown. (b) Simulated waveforms of the total current and the gap voltage.}
	\label{fig:instant}
\end{figure}

Figure \ref{fig:instant}(a) shows the waveform waveforms for the total current and the gap current of a vacuum breakdown. The electrode configuration is tip-to-plane with a gap length of 5 mm, and the voltage pulse width is 1 $\mu$s. The dashed line in Figure \ref{fig:instant}(a) represents the gap current, while the solid line represents the total current. 
By comparing the total current and the gap current, we can find a turning point, which is marked by a red dash circle, on the total current waveform at $t_0$, coinciding with the starting point of the current rise in the gap.
 
We shall prove now that the turning point in Figure \ref{fig:instant}(a) is the starting point of a vacuum breakdown.
We simulated the circuit model of Figure \ref{fig:circuit} in Simulink.
Rw was set initially to 100 M$\Omega$ and 100 ns after the application of the pulse from the voltage source dropped instantly to 1000 $\Omega$, to represent the initiation of phase P1. 
Then Rw gradually dropped to 10 $\Omega$ in the next 200 ns to simulate the ignition phase P1 of the breakdown.
As a result, we obtain the current and voltage waveforms shown in Figure \ref{fig:instant}(b).
The voltage and current waveforms are very similar to the experimental ones.
Furthermore, a turning point marked by a dash circle appears exactly at 100 ns in Figure 1, indicating that the point is a result of the breakdown in the vacuum gap. Therefore, we can confirm that indeed the time point $t_0$ is the beginning of a vacuum breakdown.
Finally, we note that the voltage keeps increasing after the breakdown instant, both in the experimental and the simulations curves.

\subsection{Verification for the repeatability of the experiments}
\begin{figure}[htbp]
   \centering
   \includegraphics[width=0.8\linewidth]{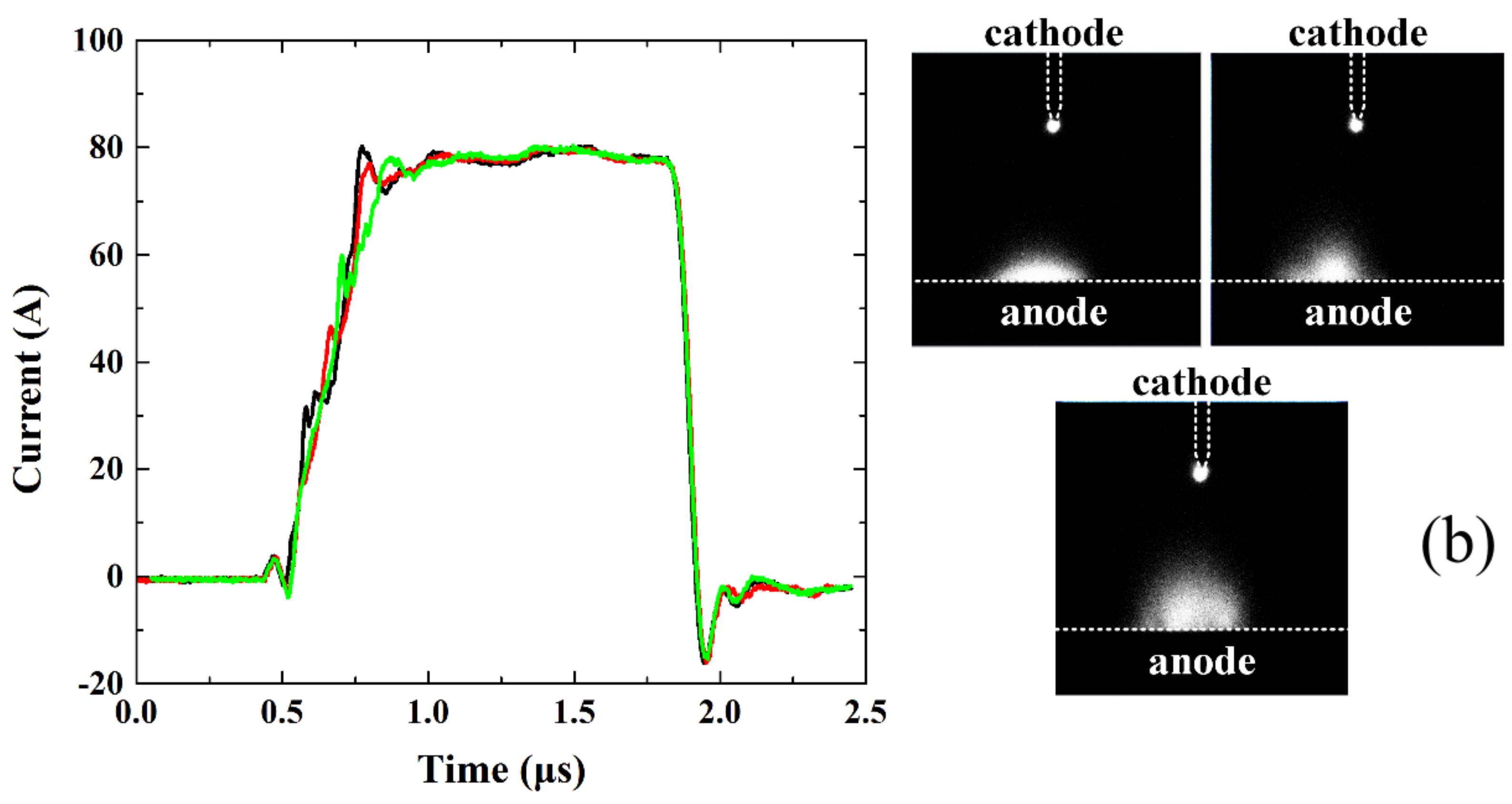}
	\caption{Repeatability test. Three independent, but identical experiments result in the same waveforms of the gap current (a). Some variation in the shapes can be observed, however the general behavor is very similar.(b) shows the three snapshots taken in these experiments with the high-speed shutter of the ICCD camera being open for 7 $\mu s$ (entire arc).}
	\label{fig:repeatability}
\end{figure}

The high voltage used in our experiments ensured the repeatability of the process with high accuracy. However, we verified this repeatability in three independent, but identical, experiments by comparing the current waveforms. We observed some variations in the shape of the waveforms; yet, the overall behaviour was found to be very similar (see Fig. \ref{fig:repeatability}(a)). In Fig. \ref{fig:repeatability}(b), we show the snapshots taken in these experiments, with an exposure time of 7 $\mu$s, that covers the whole breakdown process. All three snapshots look very similar displaying the same overall behaviour of the arc.

\subsection{Vacuum arc evolution for different pulse durations}

\begin{figure}[htbp]
	\centering
	\includegraphics[width=.6\linewidth]{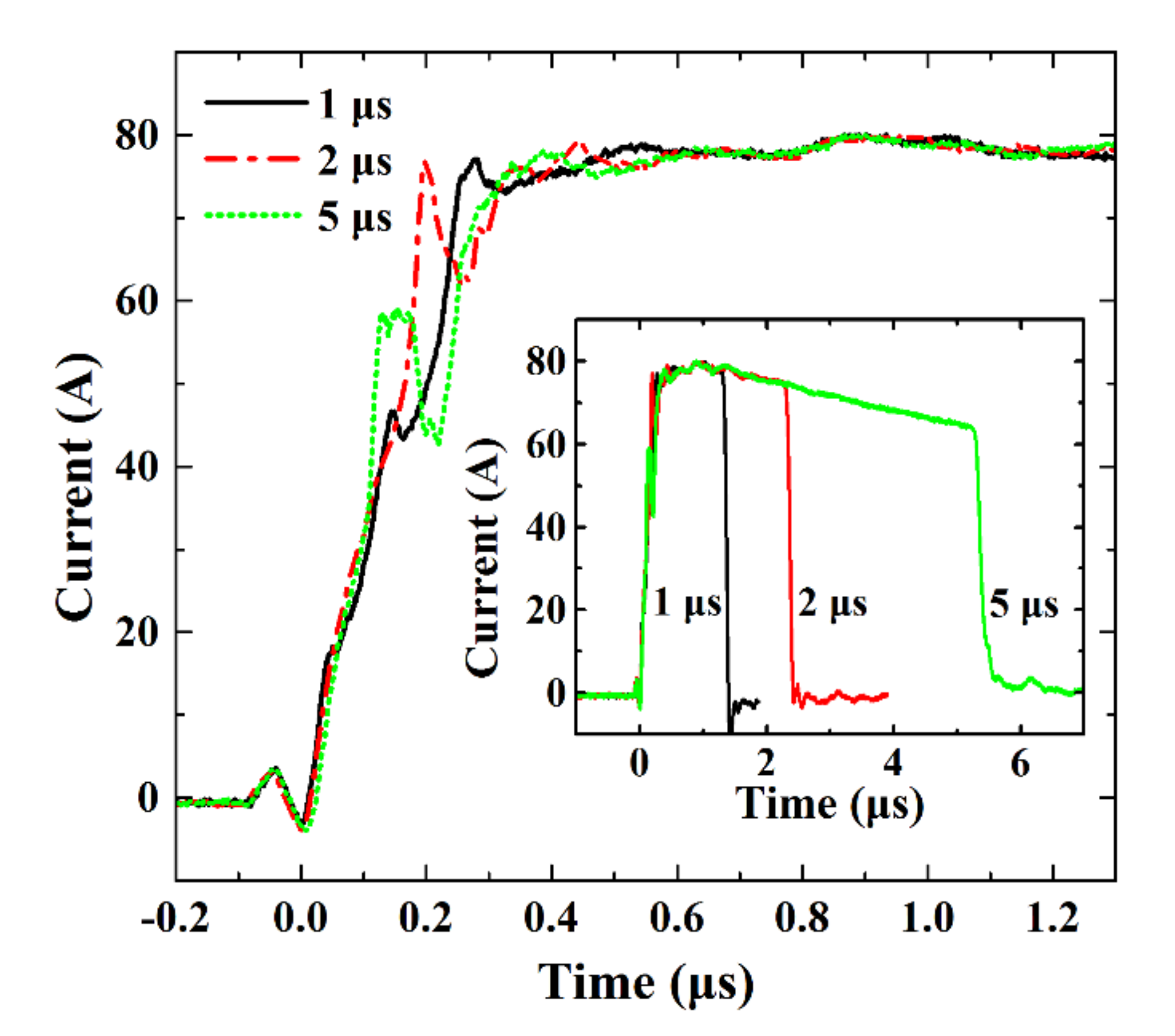}
	\caption{Current waveforms of vacuum breakdowns under different voltage pulse durations from 1 $\mu$s to 5 $\mu$s. The gap length is 5 mm.}
	\label{fig:duration}
\end{figure}

Figure \ref{fig:duration} demonstrates the current waveforms obtained for different voltage pulse durations, varying from 1 to 5 $\mu s$. 
The main figure demonstrates the rising edge of the currents focusing in phases P0 and P1, while the inset shows the corresponding complete current waveforms. 
The current pulse widths in the figure are consistent with the applied voltage pulse widths.
Furthermore, the rising dynamics of all current waveforms are very similar. 
This allows us to conclude that the initial process of a vacuum breakdown is not affected by the voltage pulse width.


\subsection{Quantitative analysis of the total gap light}

\begin{figure}[htbp]
	\centering
	\includegraphics[width=\linewidth]{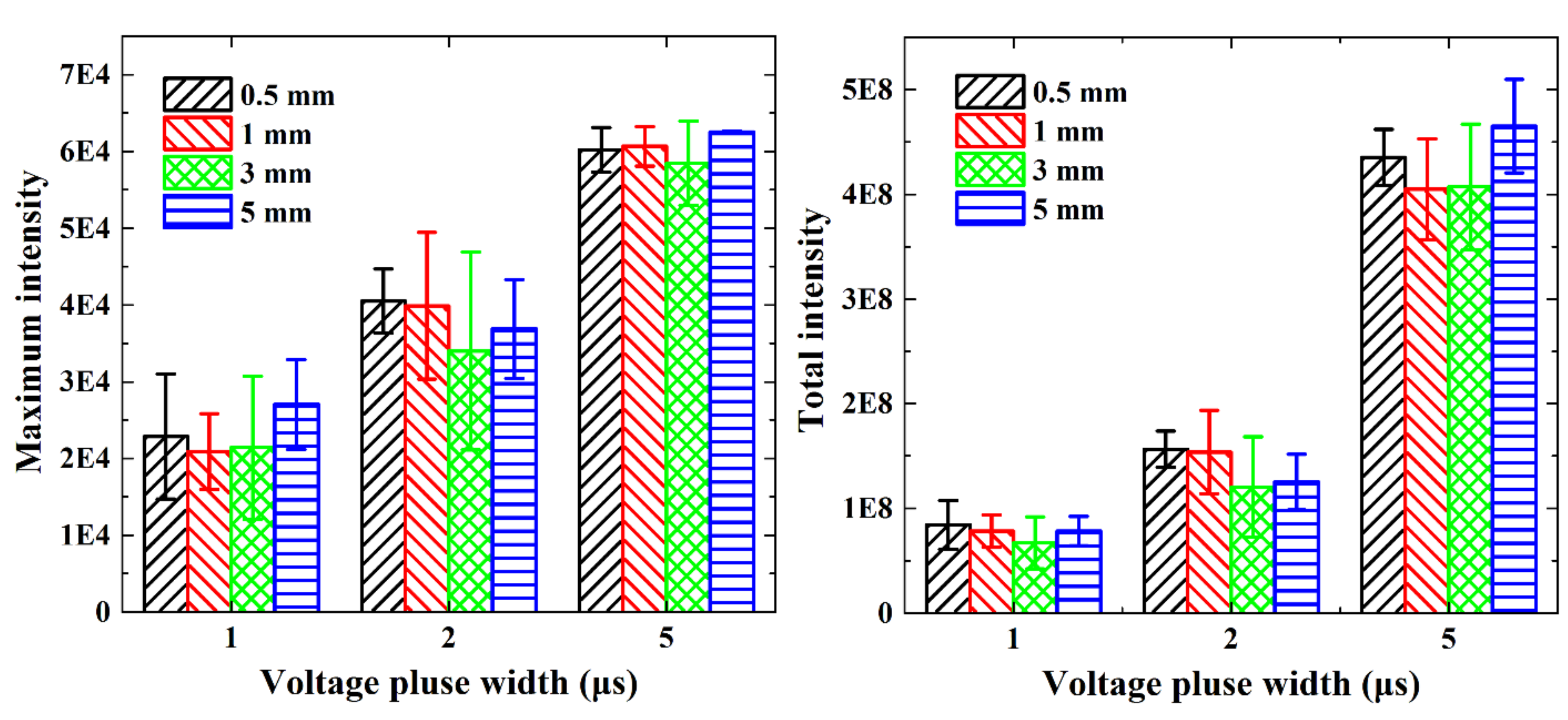}
	\caption{The maximum (a) and integrated total (b) intensity of the cathode light as a function of the voltage pulse duration and the gap length.}
	\label{fig:intensity}
\end{figure}

Figure \ref{fig:intensity} depicts the intensity of the cathode light emission for lengths varying from 0.5 mm to 5 mm and voltage pulse durations from 1 $\mu$s to 5 $\mu$s. The data were obtained by repeating the experiment 10 times for each case and calculating the corresponding average and the standard deviation for the error bars. 
Figure \ref{fig:intensity}(a) depicts the value of the maximum intensity and \ref{fig:intensity}(b) the value of the total integrated intensity, obtained by summing all the pixels of an image such as Figure 5(a) of the main text, i.e. with the camera focused at the cathode region.
Table \ref{tab:FWHM} gives the corresponding full-width-half maximum range of the peak shown in figure 5b of the main text.

\begin{table}[htbp]
\centering
\begin{tabular}{|c|c|c|}
\hline
Gap length & FWHM mean value & FWHM std. deviation\\
\hline
5 mm & 0.112 mm & 0.011 mm \\
\hline
3 mm & 0.115 mm & 0.022 mm\\
\hline
1 mm & 0.102 mm & 0.022 mm\\
\hline
0.5mm & 0.094 mm & 0.019 mm\\
\hline
\end{tabular}
\caption{\label{tab:FWHM} Full width at the half maximum of the cathode spot peak for various gap lengths. Both the mean value and the standard deviation as obtained by 10 experimental repetitions are given.}
\end{table}

As shown in the figure, both the maximum and the total intensities increase with the voltage pulse width, and the total intensities increase with the voltage pulse width in a roughly linear manner. 
For a specific voltage pulse width, however, they do not vary significantly.
The above mean that the intensity of the cathode spot remains practically constant both for any gap length or pulse duration.
The signal captured in the camera increases with the pulse duration only because the exposure time increases respectively.

\subsection{Space charge calculations}

Any type of electron emission – field, thermionic or mixed – is limited by the space charge. 
In other words, if the emitted current density exceeds a certain limit, the escape rate of the electrons from the surface become comparable to their cathode-anode flight time. The forming cloud of emitted electrons creates a significant space charge in the vacuum above the emitting surface that screens the effect of applied field. In turn, this naturally limits the current density. 
This phenomenon, known as a space charge can give an accurate estimate of the maximum current density $J_c$, which is possible for a specific geometry and experimental condition. 
For this, we use the Child-Langmuir law \cite{Child_SC}
\begin{equation} \label{eq:CL}
	J_c = \frac{4}{9 \kappa} \frac{F^2}{V^{1/2}}, \kappa = \frac{m^{1/2}}{\epsilon_0 (2e)^{1/2}}
\end{equation}
where $V$ is the applied voltage and $F$ is the magnitude of the local field, calculated without taking into account the space charge.
We note that the Child-Langmuir law is developed for planar geometries, however particle-in-cell simulations have shown that it is a valid approximation for surfaces that have radii of curvature of the order of several $\mu$m \cite{Uimanov2011}.
Here we consider only the magnitude of the field and of the current density, since their directions at least in the vicinity to the surface coincide and are always perpendicular to the emitting surface.

At every point $r$ at the cathode, $F$ is proportional to $V$, i.e. $F(r) = \xi(r) V$. 
In the latter relation, $\xi$ is the local voltage-to-field conversion factor, which is measured in units [1/m]. Substituting to eq. (\ref{eq:CL}) yields:
\begin{equation} \label{eq:Jcl}
	J_c(r) = \frac{4}{9 \kappa} \xi^2(r)V^{3/2}
\end{equation}
Given the distribution of $\xi(r)$ on the emitting cathode surface S, the maximum space-charge-limited total current can be obtained as a function of the applied voltage by integrating the current density over the emitting area S:
\begin{equation} \label{eq:Icl}
	I_c = \int_S J_c dA = \frac{4}{9 \kappa} V^{3/2} \int_S \xi^2 dA = \frac{4}{9 \kappa} V^{3/2} \zeta
\end{equation}
where $\zeta$ is a dimensionless parameter that depends on the geometry. 
In the above equation, we consider the current density to be perpendicular to the surface and therefore the vector surface integral simplifies to a scalar one. 
In the final formula giving $I_c$, $\zeta$ represents the total contribution of the geometry. 
To give an intuitive understanding of its value, if our cathode was planar, it would be $\zeta = A / d^2$ where A would be the total area of the cathode electrode.

The distribution $\xi(r)$ is calculated according to the finite element method described in the method section of the main text.
Table \ref{tab:zetas} contains the calculated maximum apex values of $\xi$ and the corresponding values of $\zeta$ as obtained by numerically integrating eq. (\ref{eq:Icl}). 
 
\begin{table}[htbp]
\centering
\begin{tabular}{|c|c|c|}
\hline
$d$  & $\zeta$ & $\xi_{max}$ \\
\hline
5 mm & 2.25 & 10.9 mm$^{-1}$ \\
\hline
3 mm & 2.65 & 11.7 mm$^{-1}$ \\
\hline
1 mm & 3.71 & 13.9 mm$^{-1}$ \\
\hline
0.5mm & 5.16 & 16.4 mm$^{-1}$ \\
\hline
\end{tabular}
\caption{\label{tab:zetas} Calculated $\zeta$ and $\xi_{max}$ for various gap distances.}
\end{table}

\begin{figure}[htbp]
	\centering
	\includegraphics[width=.6\linewidth]{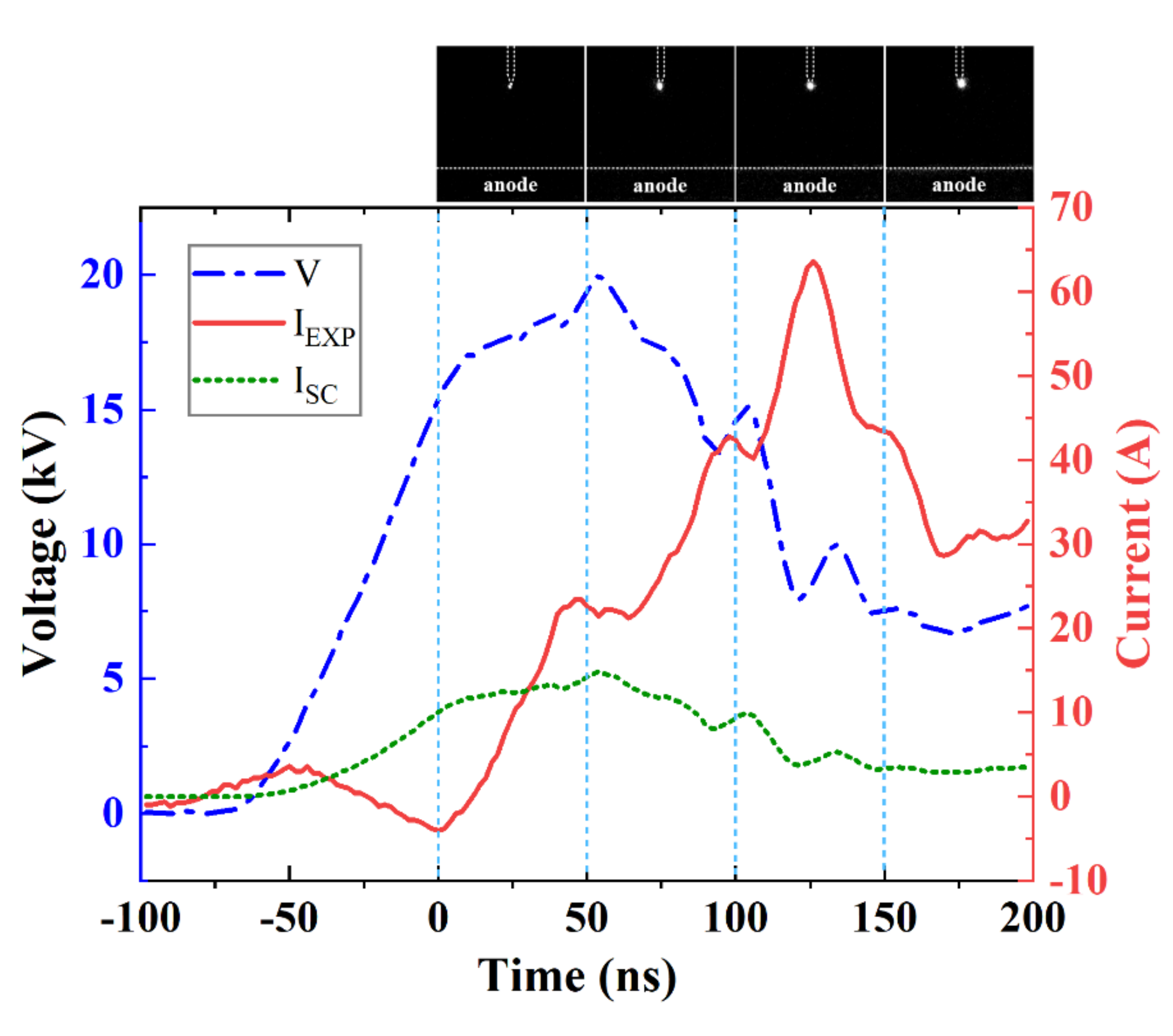}
	\caption{Voltage (blue line, left axis) and gap current (red line, right axis) waveforms measured in the default tip-to-plane configuration with the 5 mm gap length. Green dash line shows the calculated maximum space-charge limited gap current. $I_{sc}(t)$ was calculated assuming the experimental $V(t)$ waveform.}
	\label{fig:space_charge}
\end{figure}

In Figure \ref{fig:space_charge} we show the voltage waveform $V(t)$ (the left ordinate) through the gap with the standard tip-to-plane electrode configuration and the gap length 5 mm. 
The measured current waveform through the gap $I(t)$ is shown in the red curve (right ordinate), and the corresponding theoretical estimation of the maximum space-charge limited current $I_c(t)$ obtained by eq. (\ref{eq:Icl}), assuming the same $V(t)$ as in the experiment is shown in the green curve.
The camera images taken at the 50 ns intervals are placed at the top of the graph and matched with the corresponding intervals in the horizontal axis.

We clearly see that the measured current significantly exceeds the space-charge limit long before the anodic glow begins. 
We observe the same feature for all gap distances 1 mm, 3 mm, 5 mm. 
The high values of the measured current suggest that these are rather discharge currents, and not pure electron emission currents. 
During the discharge, the forming plasma compensates the vacuum space charge, and the currents extracted from the surface rise much higher than those due to a direct electron emission process.

\subsection{Additional arc image sequences}

\begin{figure}[htbp]
	\centering
	\includegraphics[width=.95\linewidth]{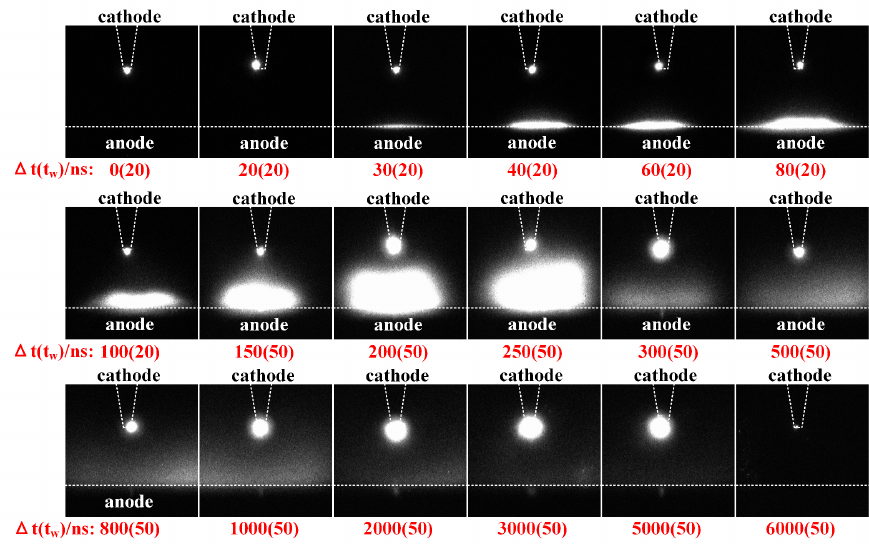}
	\caption{Arc image sequences in the gap as captured by the ICCD camera with various delay times for a gap distance of 1mm.}
	\label{fig:seq1}
\end{figure}

\begin{figure}[htbp]
	\centering
	\includegraphics[width=.95\linewidth]{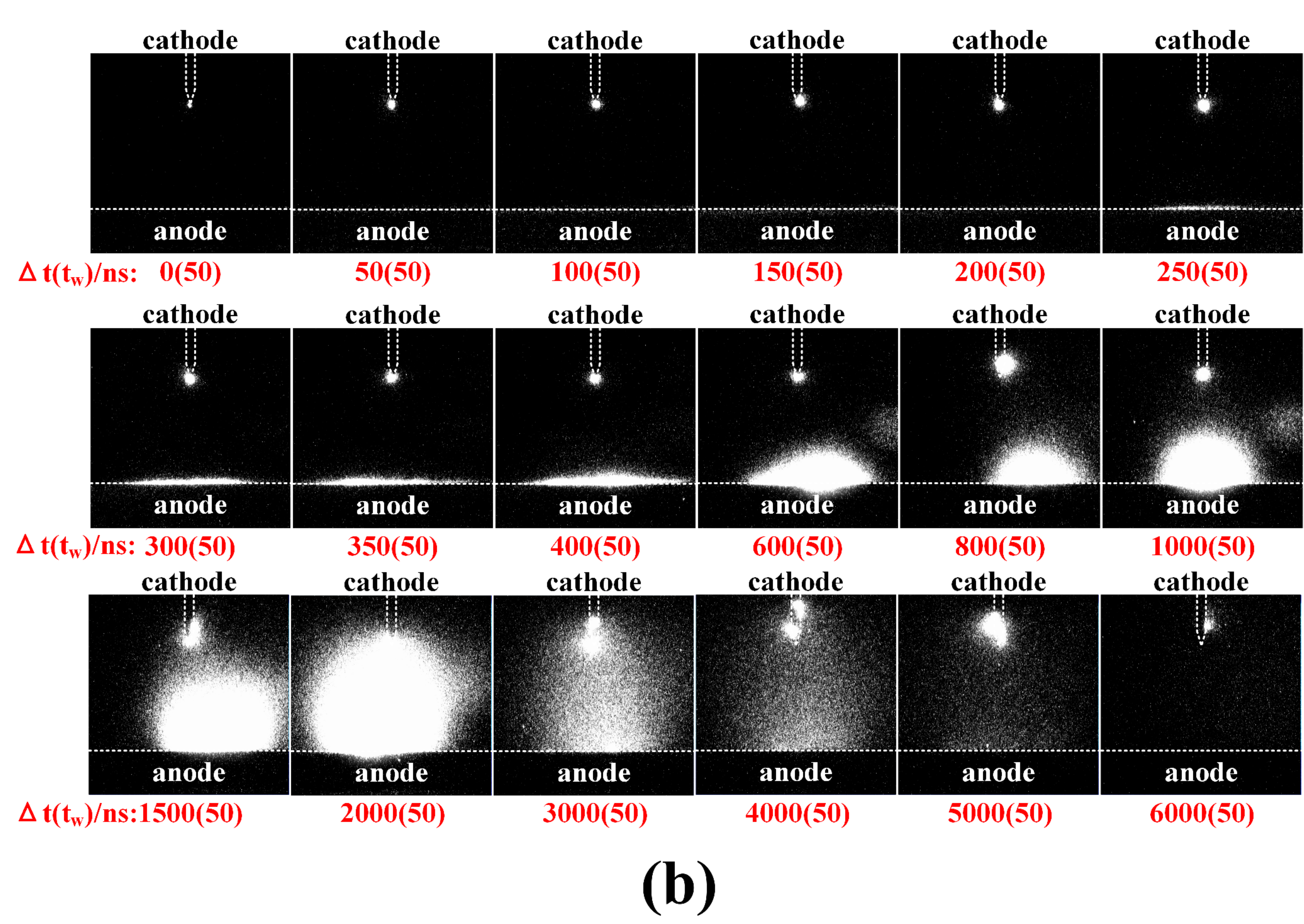}
	\caption{Arc image sequences for a gap distance of 5mm.}
	\label{fig:seq5}
\end{figure}

Figures \ref{fig:seq1} and \ref{fig:seq5} show the captured image sequences for gap distances of 1 and 5 mm respectively.

\newpage

\bibliography{bibliography/bibliography}

\end{document}